\documentclass{article}

\usepackage{amsmath,amssymb,bm}
\usepackage{url}
\usepackage{wasysym}

\usepackage[a4paper,left=2.1cm,right=2.1cm,top=2.2cm,bottom=2.2cm]{geometry}

\usepackage{natbib}
\bibpunct{[}{]}{;}{a}{,}{,}

\usepackage{graphicx,color,overpic,psfrag}
\graphicspath{{Figures/}}
\newlength\figwidth
\newtheorem{theorem}{Theorem}
\newenvironment{proof}{\noindent\textit{Proof}: }{\hfill$\blacksquare$\vskip 0.5\baselineskip}

\begin{document}

\title{Return-Map Cryptanalysis Revisited%
\thanks{%
This paper has been published in \textit{International Journal of
Bifurcation and Chaos}, vol. 16, no. 5, pp. 1157-1168, 2006. Shujun
Li is the corresponding author, contact him via his personal web
site: \texttt{http://www.hooklee.com}.}}
\author{Shujun Li\textsuperscript{1}, Guanrong
Chen\textsuperscript{2} and Gonzalo \'{A}lvarez\textsuperscript{3}}
\date{\textsuperscript{1} Department of Electronic and Information Engineering, The Hong Kong Polytechnic
University, Hung Hom, Kowloon, Hong Kong SAR, China\\[0.5em]
\textsuperscript{2} Department of Electronic Engineering, City
University of Hong Kong, 83 Tat Chee Avenue, Kowloon Tong, Hong Kong
SAR, China\\[0.5em]
\textsuperscript{3} Instituto de F\'{\i}sica Aplicada, Consejo
Superior de Investigaciones Cient\'{\i}ficas, Serrano 144---28006,
Madrid, Spain}

\maketitle

\begin{abstract}
As a powerful cryptanalysis tool, the method of return-map attacks
can be used to extract secret messages masked by chaos in secure
communication schemes. Recently, a simple defensive mechanism was
presented to enhance the security of chaotic parameter modulation
schemes against return-map attacks. Two techniques are combined in
the proposed defensive mechanism: multistep parameter modulation
and alternative driving of two different transmitter variables.
This paper re-studies the security of this proposed defensive
mechanism against return-map attacks, and points out that the
security was much over-estimated in the original publication for
both ciphertext-only attack and known/chosen-plaintext attacks. It
is found that a deterministic relationship exists between the
shape of the return map and the modulated parameter, and that such
a relationship can be used to dramatically enhance return-map
attacks thereby making them quite easy to break the defensive
mechanism.
\end{abstract}

\section{Introduction}

In the past two decades, chaotic systems have been extensively used
to construct cryptosystems in either analog
\citep{Alvarez:Survey:ICCST99, Yang:Survey:IJCC2004} or digital
\citep{ShujunLi:Dissertation2003} forms. Most analog implementations
are secure communication systems based on synchronization of the
sender and the receiver chaotic systems \citep{Pecora:CS:PRL90},
where the signal is transmitted over a public channel from the
sender to drive the receiver for achieving synchronization and
message decryption. Some different encryption structures have been
proposed: chaotic masking \citep{Kocarev:CM:IJBC92, Murali:CM:PRE94,
Feki:CM:PLA99}, chaotic switching or chaotic shift keying (CSK)
\citep{Parlitz:CSK:LIJBC92, Dedieu:CSK:IEEETCASII93,
Parlitz:DCSK:PLA94}, chaotic modulation \citep{WuChua:CDM:IJBC93,
Yang:CPM:IEEETCASI96, Parlitz:CDM:PRE96}, and the inverse system
approach \citep{Feldmann:ISA:IJCTA96}. At the same time, many
different cryptanalysis methods have also been developed to break
the proposed chaos-based secure communication systems: return-map
attacks \citep{Perez:ReturnMapCryptanalysis:PRL95,
Zhou:ExtractChaoticSignal:PLA97, Yang:ReturnMapCryptanalysis:PLA98,
Li:Chaos2005}, nonlinear prediction based attacks
\citep{Short:UnmaskingChaos:IJBC94,
Short:ChaoticSignalExtraction:IJBC97,
ZhouLai:ChaoticCryptanalysis:PRE99b}, spectral analysis attacks
\citep{Yang:SpectralCryptanalysis:PLA98, Alvarez-Li2004a},
generalized synchronization (GS) based attacks
\citep{Yang:GSCryptanalysis:IEEETCASI97,AlvarezLi:CSF2005,
Alvarez:BreakingCPM:CSF2004}, short-time period based attacks
\citep{Yang:STZCR:IJCTA95, Alvarez-Li2004b}, parameter
identification based attacks
\citep{Kocarev:BreakingParameters:IJBC96,
TaoDu:ChaoticCrytpanalysis:IJBC2003a,
Vaidya:DecodingChaoticSuperkey:CSF2003,Alvarez:PhaseSynchronization:CHAOS2004},
and so on.

Given the existence of so many different attacks, it has become a
real challenge to design highly secure chaos-based communication
systems against all known attacks. Three general countermeasures
have been proposed in the literature: 1) using more complex
dynamical systems, such as high-dimensional hyperchaotic systems
or multiple cascaded (heterogeneous) chaotic systems
\citep{Grassi:3CC:IJCTA99,
Murali:HeterogeneousChaosCryptography:PLA2000,
Yao:HyperChaos:DCDISB2003}; 2) introducing traditional ciphers
into the chaotic cryptosystems
\citep{Yang-Wu-Chua:ChaoticCryptography:IEEETCASI1997,
Grassi:3CC:IEEETCASI99, Lian:Fuzzy3CC:IJBC2003}; 3) introducing an
impulsive (also named sporadic) driving signal instead of a
continuous signal to realize synchronization
\citep{Yang-Chua:ImpulsiveChaos:IJBC1997, ZYHe:SCCS:IEEETCASI2000,
Khadra:IS:DCDISB2003}. The first countermeasure has been found
insecure against some attacks
\citep{Short:UnmaskingHyperchaos:PRE98,
ZhouLai:ChaoticCryptanalysis:PRE99b,
Huang:UnmaskingChaosWavlet:IJBC2001,
TaoDu:ChaoticCrytpanalysis:IJBC2003a}, and some security defects
of the second countermeasure have also been reported
\citep{Short:ChaoticCrptanalysis:IEEETCASI2001}, but the last one
has not yet been cryptanalyzed to date.

Besides the above three general countermeasures, there also exist
some specially-designed countermeasures that can be used to resist
certain attacks. This paper studies two such countermeasures,
recently proposed by \cite{Indian:MSCPM:IJBC2001}, against
return-map attacks. These two proposed countermeasures are
multistep parameter modulation and alternative driving of
transmitter variables, which have been combined to construct a new
secure communication scheme for binary signal transmissions. After
refining return-map attacks via a deterministic relationship
between the return map and a parameter $b_s$, we found that the
security of the first countermeasure was much over-estimated in
\citep{Indian:MSCPM:IJBC2001}, and that the combination of the two
countermeasures can be easily separated in some way so as to
disable the second countermeasure. The aforementioned
deterministic relationship between the return map and the
parameter $b_s$ is reported in this paper, for the first time in
the literature, which is useful not only for engineering studies
on chaos-based secure communications but also for theoretical
studies on the dynamics of chaotic systems.

The rest of this paper is organized as follows. In the next
section, a brief introduction to return-map attacks and related
countermeasures is given. Section \ref{section:BreakingMSPM}
re-evaluates the security of the multistep parameter modulation
scheme, by exploiting a deterministic relationship between the
shape of the return map and the modulated parameter $b_s$. The
original return-map attack proposed by
\cite{Perez:ReturnMapCryptanalysis:PRL95} will be enhanced. In
Sec. \ref{section:BreakingADTV}, cryptanalysis of the scheme of
alternative driving of transmitter variables is studied in detail.
The last section concludes the paper.

\section{Return-Map Attacks and Related Countermeasures}

The return-map attack method was first proposed by
\cite{Perez:ReturnMapCryptanalysis:PRL95} to break chaotic
switching (binary parameter modulation) and chaotic masking
schemes based on the Lorenz system, which was then studied by
\cite{Yang:ReturnMapCryptanalysis:PLA98} to break chaotic masking,
switching and non-autonomous modulation schemes based on Chua's
circuit. In \citep{Zhou:ExtractChaoticSignal:PLA97}, the
return-map attack method was also used to break a DCSK scheme
based on a discrete-time chaotic map \citep{Parlitz:DCSK:PLA94}.
Without loss of generality, this paper will focus on the attack
scheme of \citeauthor{Perez:ReturnMapCryptanalysis:PRL95} on the
Lorenz system thereby demonstrating how the return map is
constructed and how the attack works to break a typical chaotic
switching scheme proposed in \citep{Cuomo:CPM_CM:PRL93}.

Consider the following Lorenz system used as the sender:
\begin{eqnarray}
\dot{x}_s & = & \sigma(y_s-x_s),\nonumber\\
\dot{y}_s & = & r_sx_s-y_s-x_sz_s,\\
\dot{z}_s & = & x_sy_s-b_sz_s,\nonumber
\end{eqnarray}
where $\sigma,b_s,r$ are system parameters, and the value of $b_s$
is modulated by $m(t)$, the digital plain-signal for secure
transmission, as follows:
\[
b_s=\begin{cases}
b_0, & m(t)=0,\\
b_1, & m(t)=1.
\end{cases}
\]
To transmit $m(t)$ to the receiver end, a variable of the sender
system, such as $x_s$, is sent out, which will be used to induce
synchronization of the receiver system, resulting in:
\begin{eqnarray}
\dot{x}_r & = & \sigma(y_r-x_r),\nonumber\\
\dot{y}_r & = & r_rx_s-y_r-x_sz_r,\\
\dot{z}_r & = & x_sy_r-b_rz_r,\nonumber
\end{eqnarray}
where $b_r=b_0$. When $m(t)=0$, the intended synchronization can
be reached, while when $m(t)=1$, the synchronization error always
remains at a certain finite order. Then, it is easy to decode the
secret signal $m(t)$ by checking the power energy $(x_r-x_s)^2$
with a digital filter. Following \cite{Cuomo:CPM_CM:PRL93}, the
parameters are set as $\sigma=16$, $r=45.6$, $b_0=4.0$ and
$b_1=4.4$.

\begin{figure}[!htb]
\centering
\psfrag{Am and -Cm}{$A_m$ and $-C_m$}%
\psfrag{Bm and -Dm}{$B_m$ and $-D_m$}%
\includegraphics[width=\figwidth]{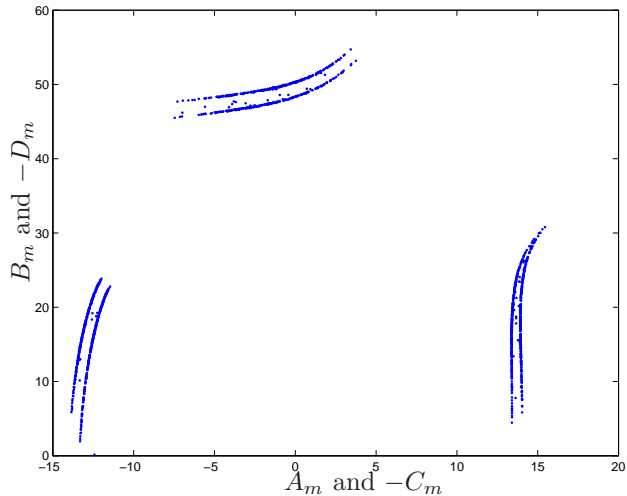}
\caption{The return maps constructed for a typical chaotic
switching scheme.}\label{figure:RM_CSK}
\end{figure}

However, the above chaotic switching scheme can be easily broken
with the return map constructed from $x_s$ as pointed out in
\citep{Perez:ReturnMapCryptanalysis:PRL95}. Assuming that $X_m$ and
$Y_m$ are the $m$-th maxima and $m$-th minima of $x_s$,
respectively, define the following four variables:
$A_m=\frac{X_m+Y_m}{2}$, $B_m=X_m-Y_m$, $C_m=\frac{X_{m+1}+Y_m}{2}$,
$D_m=Y_m-X_{m+1}$, and then construct two return maps, ($A_m$ vs
$B_m$) and ($-C_m$ vs $-D_m$), as shown in Fig. \ref{figure:RM_CSK}.
The two maps are actually equivalent to each other, so we only
consider the map ($A_m$ vs $B_m$) in this paper. Note that there are
three segments in the return map, and each segment is further split
into two strips. It is obvious that the split of the map is caused
by the switching of the value of $b_s$ between $b_0$ and $b_1$.
Thus, by checking which strip the point $(A_m,B_m)$ falls on, one
can easily unmask the current value of the digital signal $m(t)$.
Since one has to assign either 0-bit or 1-bit to a strip in each
segment, it was claimed in
\citep{Perez:ReturnMapCryptanalysis:PRL95} that there are only seven
chances to make wrong assignments, which can be easily detected by
observing the waveform of the reconstructed digital signal $m(t)$.

\psfrag{Am}{$A_m$}\psfrag{Bm}{$B_m$}

In recent years, some different countermeasures have been proposed
to resist the above return-map attack. In \citep{BuWang:CSF2004},
a periodic signal $g_0(t)=A\cos(\omega t+\phi_0)$ is combined with
$z_s$ to modulate the transmitted signal $x_s$ so as to blur the
reconstructed return map in order to frustrate the attacker.
However, soon it was broken as reported in
\citep{CheeXuBishop:CSF2004, WuHuZhang:CSF2004, Alvarez:CSF2005}
via distinguishing the parameters $\omega,\phi_0$ and removing the
modulating signal. A modified scheme of the original method of
\cite{BuWang:CSF2004} was proposed in \citep{WuHuZhang:CSF2004} to
further improve its security. Our recent work shows that this
modified modulating scheme is still not secure enough
\citep{LiAlvarez:CSF2005} and that the modulating signal can still
be effectively removed via parameters estimation.

\begin{figure}[!htb]
\centering
\psfrag{Segment 1}{Segment 1}%
\psfrag{Segment 2}{Segment2}%
\psfrag{Segment 3}{Segment 3}
\begin{minipage}{\figwidth}
\centering
\includegraphics[width=\textwidth]{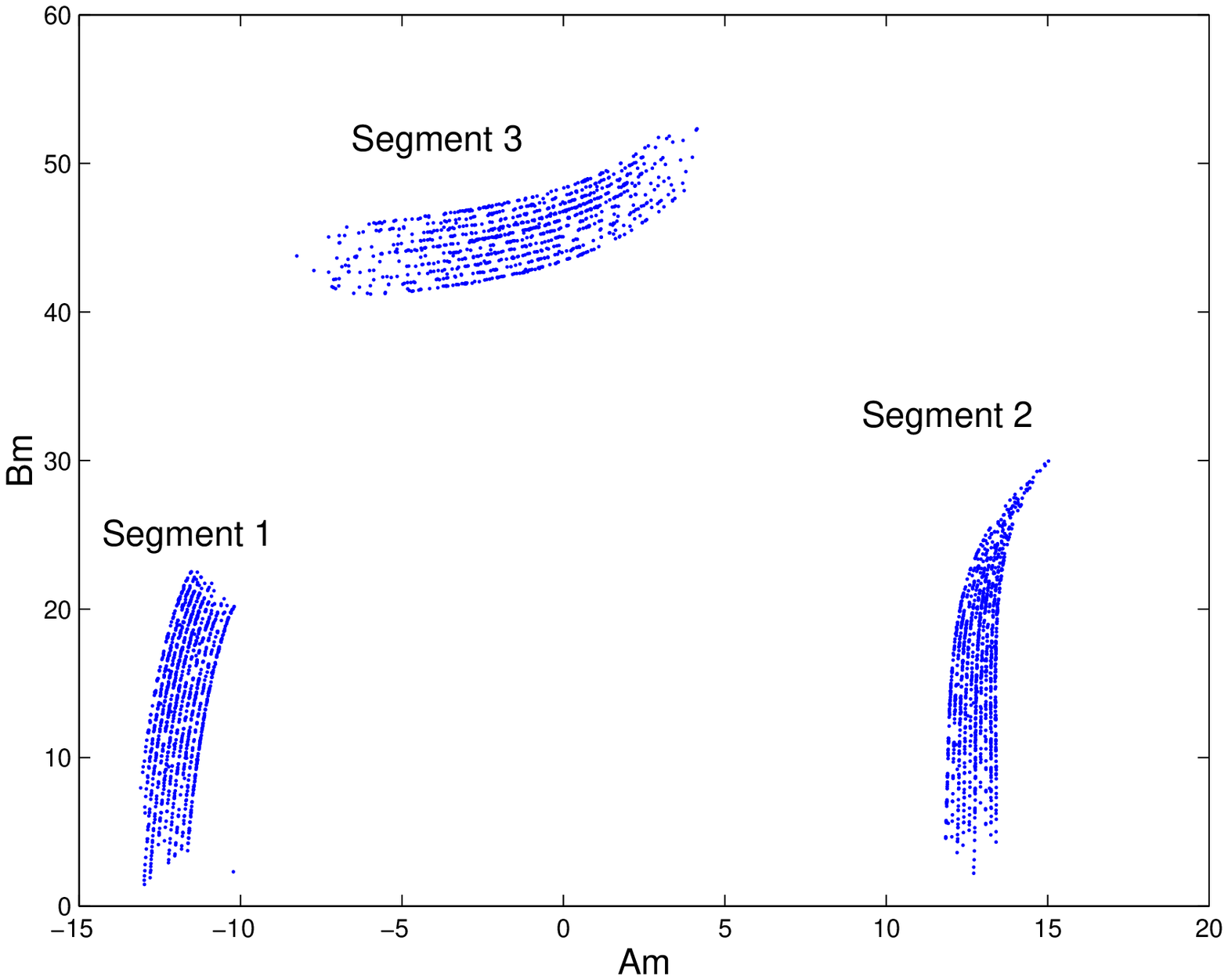}
a) a full view of the return map
\end{minipage}
\begin{minipage}{\figwidth}
\centering
\includegraphics[width=\textwidth]{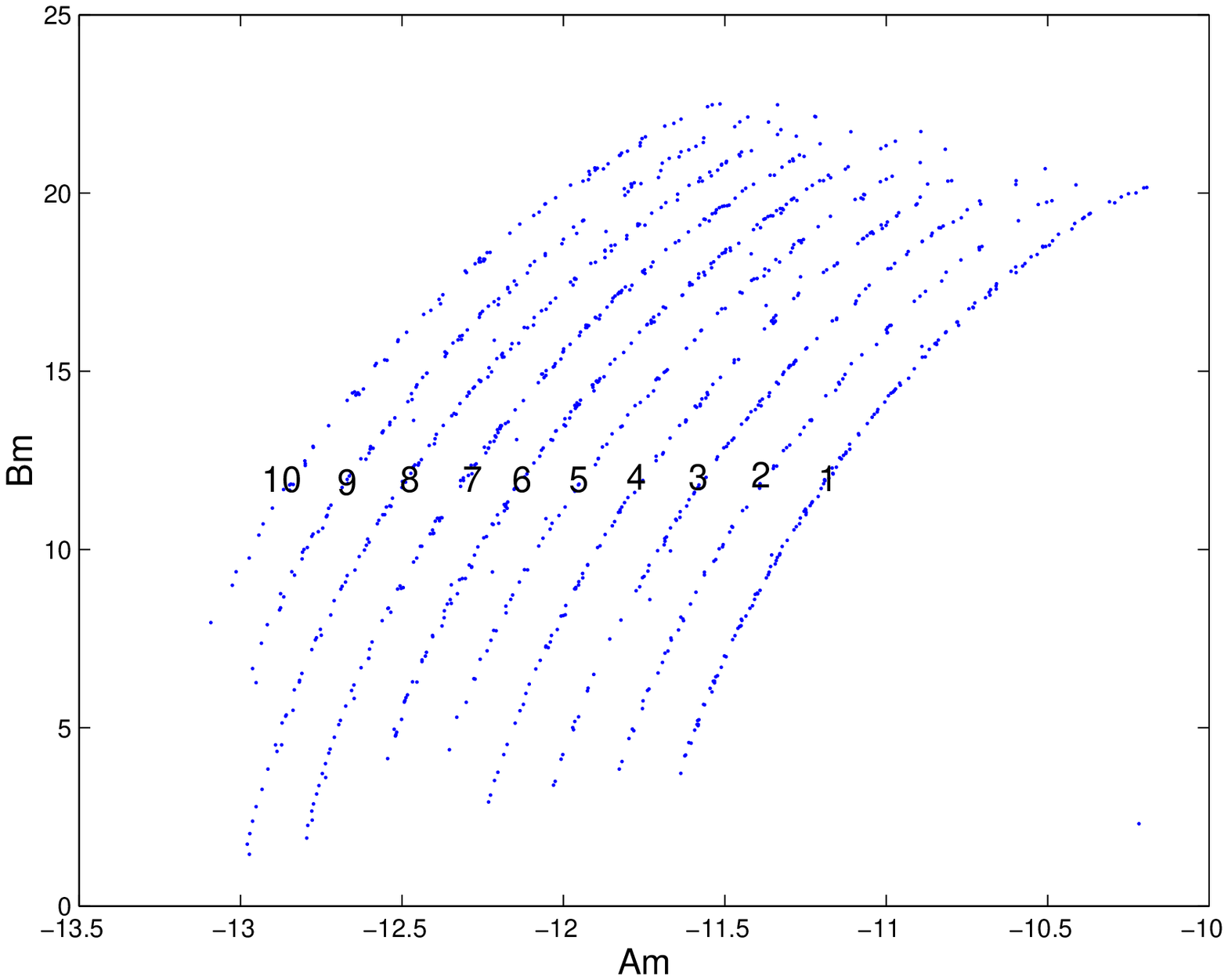}
b) a local view of Segment 1
\end{minipage}
\caption{The return map constructed from $x_s$ in multistep
parameter modulation.}\label{figure:RM_MSPM}
\end{figure}

\begin{figure}[!htb]
\centering
\begin{minipage}{\figwidth}
\centering
\includegraphics[width=\textwidth]{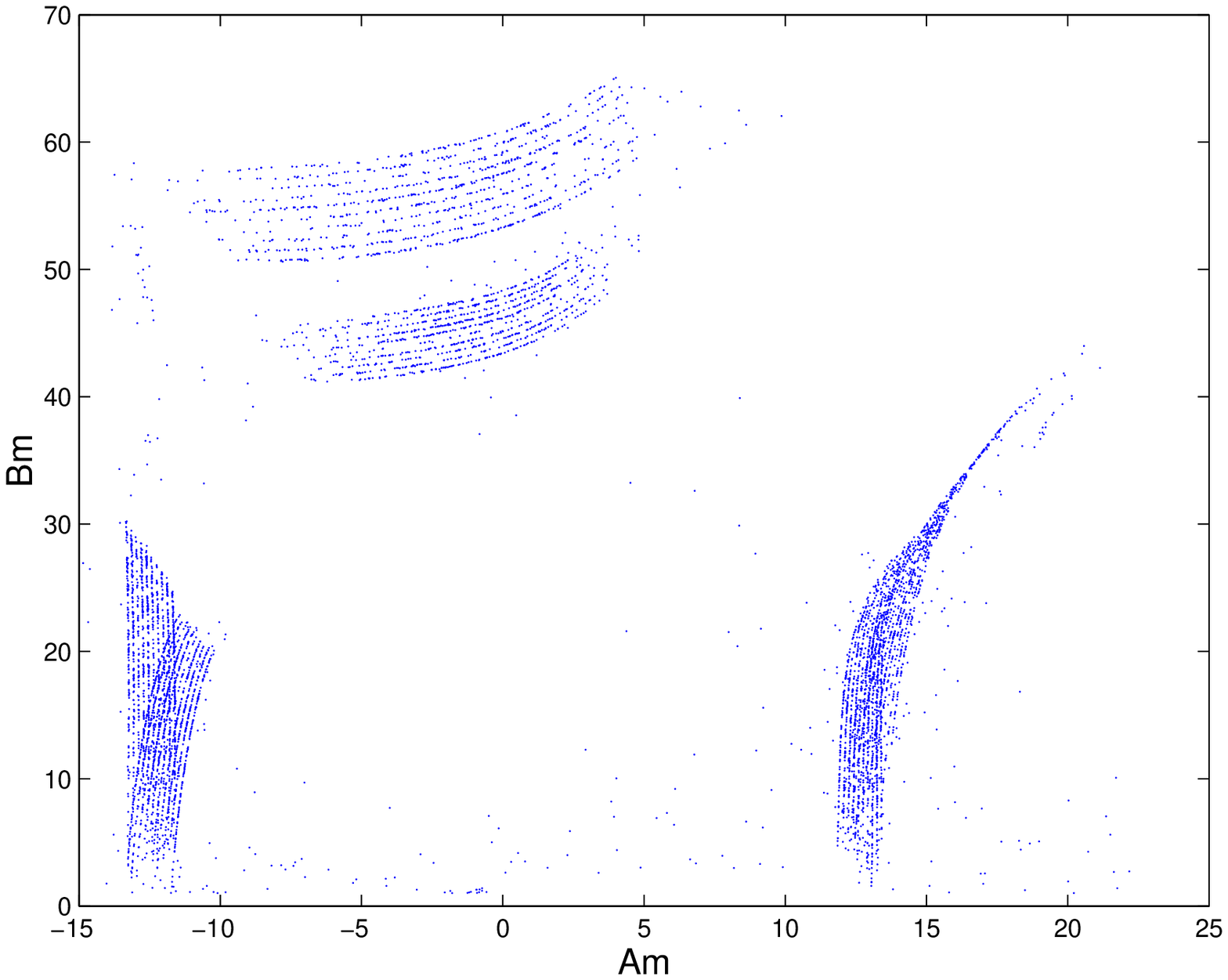}
a) a full view of the return map
\end{minipage}
\begin{minipage}{\figwidth}
\centering
\includegraphics[width=\textwidth]{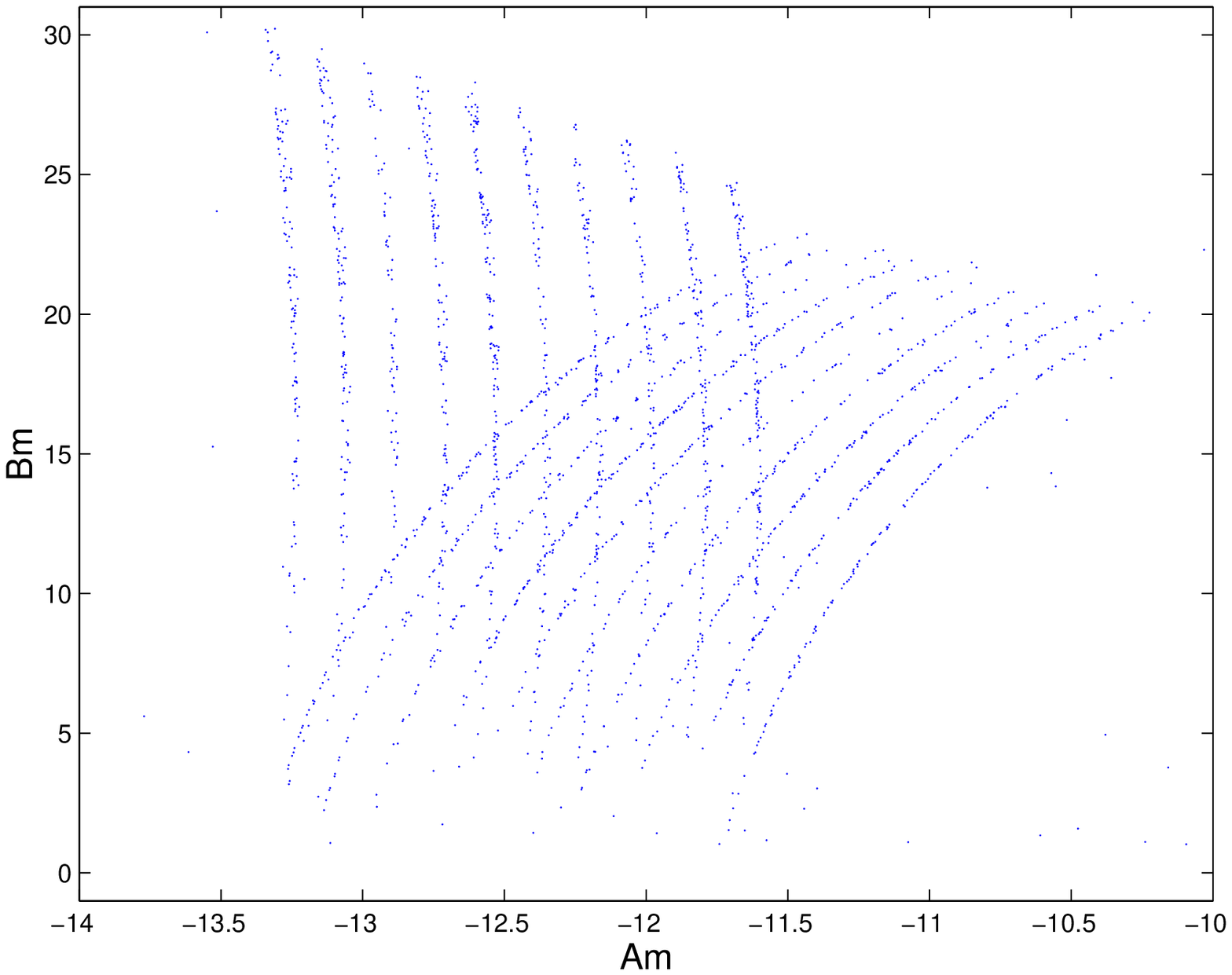}
b) a local view of Segment 1
\end{minipage}
\caption{The return map constructed in multistep parameter
modulation with alternative $x_s/y_s$
driving.}\label{figure:RM_MSPMAD}
\end{figure}

In \citep{Indian:MSCPM:IJBC2001}, two new countermeasures were
proposed and combined to enhance the security of chaotic switching
schemes against return-map attacks. The first countermeasure is to
increase the number of strips by modulating $b_s$ between $2n$
different values: $b_{0,1},\cdots,b_{0,n}$ and
$b_{1,1},\cdots,b_{1,n}$, where $b_{0,1},\cdots,b_{0,n}$
correspond to $m(t)=0$ and $b_{1,1},\cdots,b_{1,n}$ correspond to
$m(t)=1$. This countermeasure is called \textit{multistep
parameter modulation}, and accordingly the original two-valued
chaotic switching scheme is called \textit{single-step parameter
modulation}. It was claimed that the chances to make wrong
assignments become $(2^{2n}-2)^3-1\approx 2^{6n}$ and that the
security against return-map attacks is dramatically enhanced even
when $n$ is not too large. Figure \ref{figure:RM_MSPM} shows the
return map constructed from $x_s$ when the multistep parameter
modulation is used, where $n=5$ and
$b_{0,i}\in\{3.2,3.4,3.6,3.8,4.0\}$,
$b_{1,i}\in\{3.1,3.3,3.5,3.7,3.9\}$. It can be seen that each
segment is split into $2n=10$ strips. The second countermeasure is
to alternatively use $x_s$ and $y_s$ as the driving signal to
force the receiver system to synchronize with the sender, which
will further split the constructed return map into two parts: one
corresponds to the map from $x_s$ and another to the map from
$y_s$, as shown in Fig. \ref{figure:RM_MSPMAD}\footnote{Different
from $x_s$, there exist some small fluctuations in $y_s$. The
faked maxima and minima induced by the small fluctuations should
be removed from the return map; otherwise, the map will become
completely meaningless. For the return map plotted in Fig.
\ref{figure:RM_MSPMAD}, therefore, if the difference between two
consecutive maxima and minima is less than 1, they will be
omitted.}. It can be seen that two segments of the $x_s$-map and
the $y_s$-map largely overlap each other. In a multistep parameter
modulation system, the receiver contains $2n$ different driven
sub-systems, which are used to realize synchronization for the
$2n$ different values of $b_s$, respectively. When alternative
driving is also applied, the number of sub-systems is doubled to
be $4n$, among which $2n$ correspond to $x_s$-driving
synchronization and another $2n$ to $y_s$-driving synchronization.
For more details about the two countermeasures, see
\citep{Indian:MSCPM:IJBC2001}.

\section{Re-Evaluating the Security of Multistep Parameter Modulation}
\label{section:BreakingMSPM}

The security of multistep parameter modulation relies on the fact
that the attacker has to assign 0-bits or 1-bits for all strips in
the return map. Since there are $6n$ stripes in total, the success
probability to make a right assignment is $\frac{1}{2^{6n}}$,
i.e., the attack complexity is $2^{6n}$. Note that the above
analysis on the security is more rigorous, from the
cryptographical point of view, than the one given in
\citep{Indian:MSCPM:IJBC2001}, where the latter enumerated the
number of making wrong assignments under the assumption that the
first assignment is correct. Of course, the order of the estimated
attack complexity is the same.

\begin{figure}[!htb]
\centering
\begin{minipage}{\figwidth}
\centering \psfrag{bs=3}{$b_s=3$}\psfrag{bs=4}{$b_s=4$}
\includegraphics[width=\textwidth]{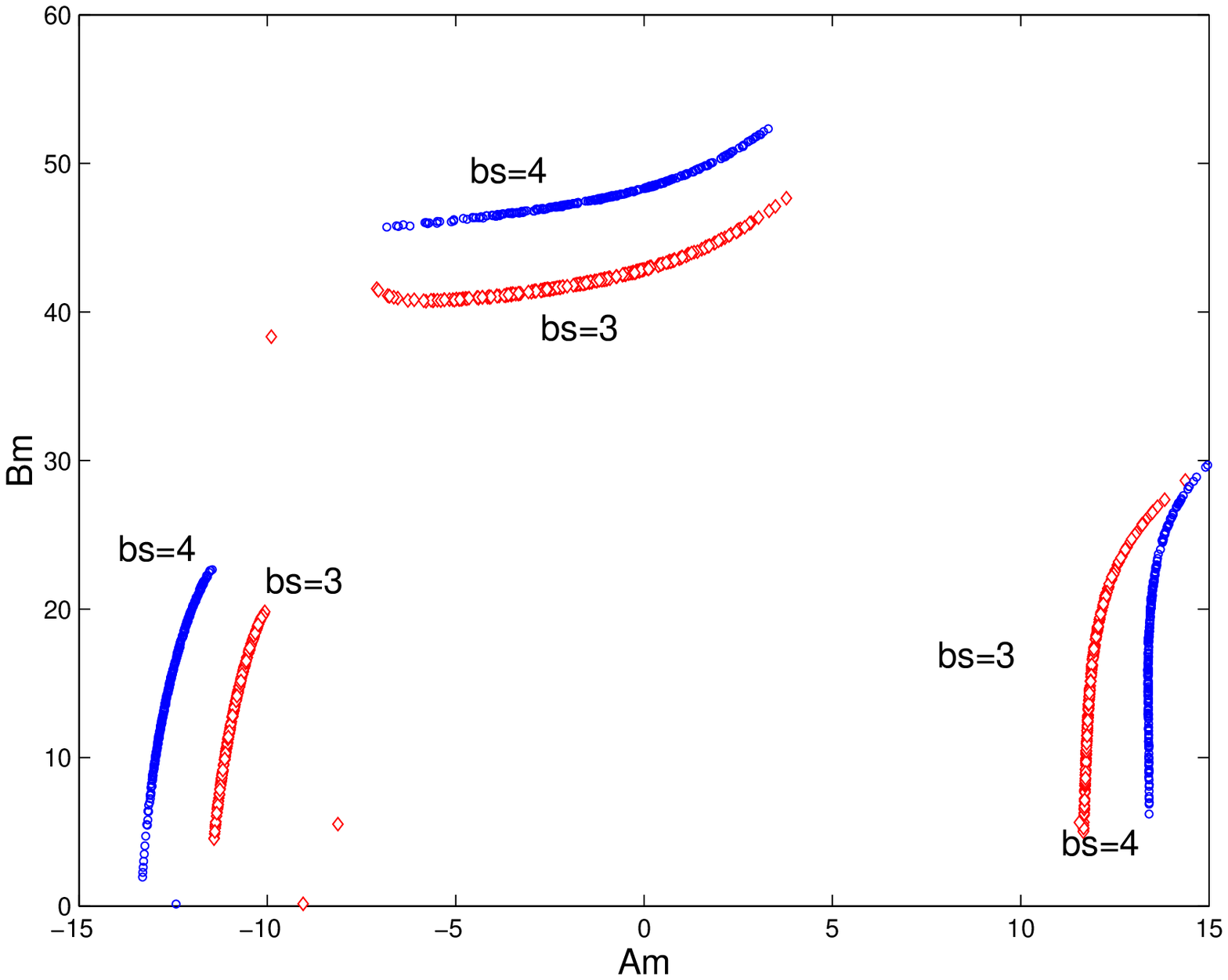}
a) the return maps corresponding to $b_s=3$ and $b_s=4$
\end{minipage}
\begin{minipage}{\figwidth}
\centering \psfrag{bs=3.1}{$b_s=3.1$}\psfrag{bs=4.0}{$b_s=4.0$}
\includegraphics[width=\textwidth]{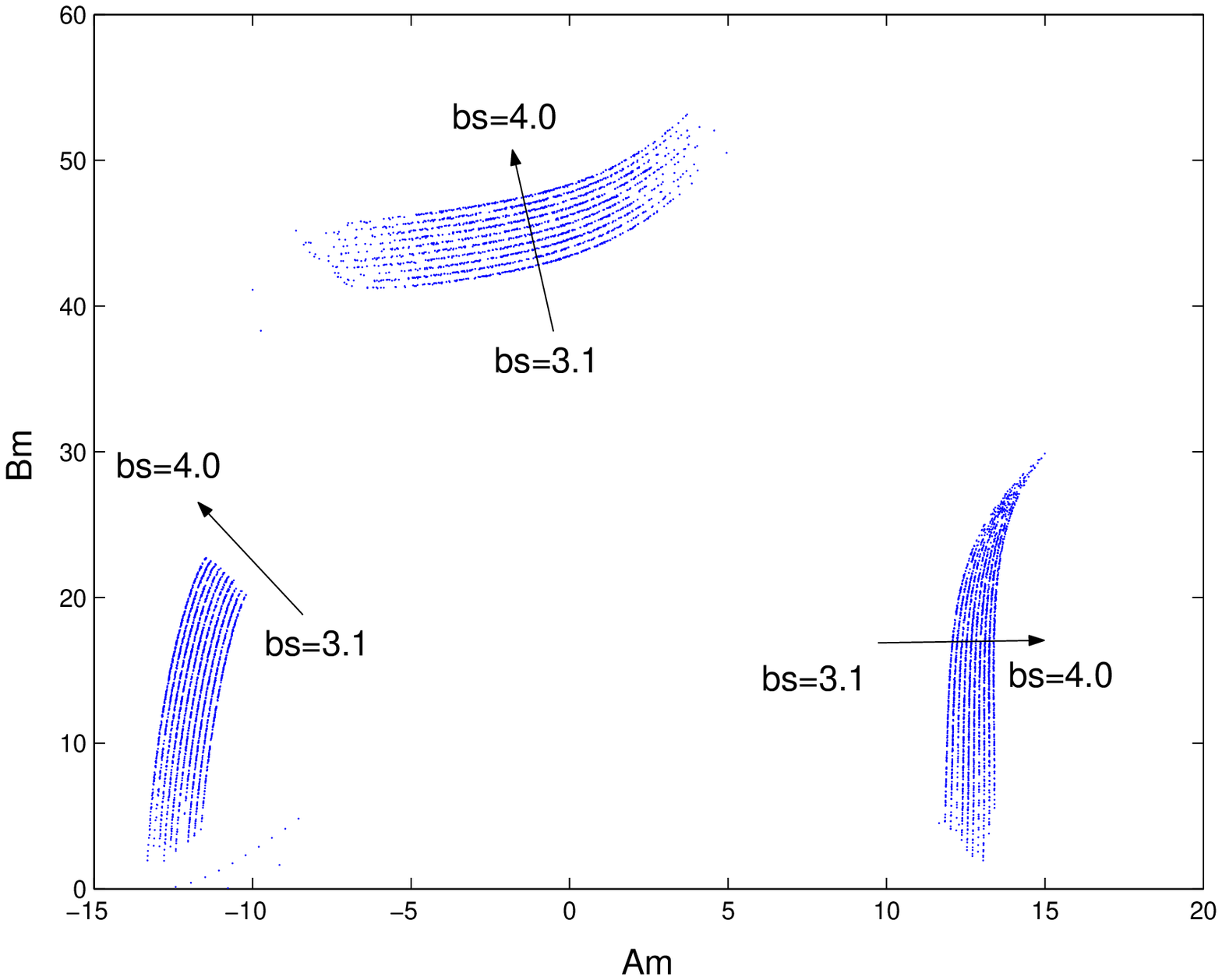}
b) the return maps corresponding to $b_s=3.1,3.2,\cdots,3.9,4.0$
\end{minipage}
\caption{A deterministic relationship between the return map and
the modulated parameter $b_s$.}\label{figure:RM_bs}
\end{figure}

The above security estimation is based on the assumption that all
$6n$ strips are independent of each other. However, we found that
this assumption is not true and that there exists a deterministic
relationship between the positions of the strips and the $2n$
different values of the modulated parameter $b_s$, and this
relationship will dramatically reduce the attack complexity in all
attacking scenarios. In Fig. \ref{figure:RM_bs}a, the two return
maps corresponding to $b_s=3$ and $b_s=4$ respectively are plotted
to show such a deterministic relationship. One can see that the
three segments corresponding to $b_s=3$ are closer to the origin,
while the three segments corresponding to $b_s=4$ are farther.
This means that there exist only two possibilities to assign the
0/1-bits to all strips in the chaotic switching scheme (see Fig.
\ref{figure:RM_CSK}): for all three segments, assign 0-bit (or
1-bit) to the strip closer to the origin and 1-bit (or 0-bit) to
the other one. If the relationship between $b_0$ and $b_1$ is also
known to the attacker, he can uniquely determine the right
assignment to completely break the plain-signal. Apparently, the
above analysis can be easily generalized to multistep parameter
modulation. Figure \ref{figure:RM_bs}b shows the return maps
corresponding to the 10 different values of $b_s$ used in
\citep{Indian:MSCPM:IJBC2001}. It can be seen that Fig.
\ref{figure:RM_bs}b is almost identical with the return map shown
in Fig. \ref{figure:RM_MSPM}a. Thus, it is easy to mark each strip
of the return map shown in Fig. \ref{figure:RM_MSPM}a with one of
the $2n=10$ possible values of $b_s$. For example, for Segment 1
shown in Fig. \ref{figure:RM_MSPM}b, the $i$-th strip corresponds
to $b_s=3.0+0.1i$. This means that the task of assigning 0/1-bits
to $6n$ strips is changed to another equivalent task of assigning
0/1-bits to $2n$ different values of $b_s$. Considering that there
are $n$ values corresponding to 0-bits and other $n$ values to
1-bits, one can easily deduce that the number of all possible bit
assignments is $2\cdot\binom{2n}{n}=2\cdot\frac{(2n)!}{(n!)^2}$,
which is $O\left(\frac{2^{2n}}{\sqrt{n}}\right)$ when $n\gg 1$
following Stirling's approximation \citep{StirlingApproximation}.
As a conclusion, the attack complexity is always much smaller than
$O\left(2^{6n}\right)$, the original complexity estimated in
\citep{Indian:MSCPM:IJBC2001}. Table \ref{table:Complexity} shows
a comparison of the two complexities. From the cryptographical
point of view, based on today's computer technology, a practically
secure cryptosystem should have a complexity of order
$O\left(2^{100}\right)$ \citep{Schneier:AppliedCryptography96},
which requires $n\geq 50$ following the data shown in Table
\ref{table:Complexity}. However, in this case, $4n\geq 200$
sub-systems have to be constructed to realize the decryption of
the transmitted digital signal $m(t)$, which makes the
implementation too costly for most practical applications. If the
security can be relaxed to order of $2^{50}$, $4n\geq 32$
sub-systems are enough to be practical in some applications
(though still much more costly than other chaos-based secure
communication systems). Note that the implementation cost will be
acceptable in practice, if all the sub-systems can be realized
with the same chaotic circuit.

\begin{table}[!htb]
\centering \renewcommand\arraystretch{1.5} \caption{A comparison of
the real complexity $2\cdot\binom{2n}{n}$ and the over-estimated
complexity $2^{6n}$.}\label{table:Complexity}
\begin{tabular}{c||*{12}{c|}c}
\hline $n$ & 8 & 10 & 12 & 14 & 16 & 18 & 20 & 25 & 30 & 35 & 40 & 45 & 50\\
\hline $2\cdot\binom{2n}{n}\approx$ & $2^{14.7}$ & $2^{18.5}$ &
$2^{22.4}$ & $2^{26.3}$ & $2^{30.2}$ & $2^{34.1}$ & $2^{38}$ &
$2^{47.8}$ & $2^{57.7}$ & $2^{67.6}$ & $2^{77.5}$ & $2^{87.4}$ & $2^{97.3}$\\
\hline $2^{6n}$ & $2^{48}$ & $2^{60}$ & $2^{72}$ & $2^{84}$ &
$2^{96}$ & $2^{108}$ & $2^{120}$ & $2^{150}$ & $2^{180}$ & $2^{210}$
& $2^{240}$ & $2^{270}$ & $2^{300}$\\\hline
\end{tabular}
\end{table}

Note that one can extract some right bits even with a wrong bit
assignment. For instance, for the example given in
\citep{Indian:MSCPM:IJBC2001}, 1-bits are assigned to
$b_s\in\{3.1,3.3,3.5,3.7,3.9\}$ and 0-bits to
$b_s\in\{3.2,3.4,3.6,3.8,4.0\}$, so one can get about 80\% of right
bits with the following bit assignment: 1-bits are assigned to
$b_s\in\{3.1,3.3,3.5,3.7,\bm{4.0}\}$, and 0-bits to
$b_s\in\{3.2,3.4,3.6,3.8,\bm{3.9}\}$, where the bold values
correspond to wrong bits. Generally speaking, if there are $2i$
values corresponding to wrong bits, the bit error ratio (BER) at the
attacker end will be $i/n$, i.e., the probability to get right bits
is $1-(i/n)$. Note that when $i<n/2$, the attacker can simply flip
all assigned bits to get a lower BER $(n-i)/n=1-(i/n)$. From such a
point of view, the worst bit assignment occurs when $i=\lfloor
n/2\rfloor$ or $\lceil n/2\rceil$. Considering that the bit
assignment can be regarded as an equivalent of the secret key, the
above fact means that the decryption of multistep parameter
modulation is insensitive to the mismatch of the secret key.
However, such an insensitivity does not reduce the attack complexity
by too much, since the number of wrong assignments corresponding to
$i=\lfloor n/2\rfloor$ or $\lceil n/2\rceil$ is in the same order as
the complexity $O\left(\frac{2^{2n}}{\sqrt{n}}\right)$ when $n\gg
1$: the number is $2\cdot\binom{n}{\lfloor
n/2\rfloor}\cdot\binom{n}{n-\lfloor
n/2\rfloor}=2\cdot\binom{n}{\lfloor n/2\rfloor}\cdot\binom{n}{\lceil
n/2\rceil}\approx O\left(\frac{2^{2n}}{n}\right)$, which is not much
smaller than $O\left(\frac{2^{2n}}{\sqrt{n}}\right)$.

In cryptography, there are many different attacking scenarios
\citep{Schneier:AppliedCryptography96}. A cryptographically secure
cryptosystem should be immune to all kinds of attacks. The above
attack complexity of multistep parameter modulation is for the
simplest attack -- the ciphertext-only attack, where the attacker
can only observe some ciphertexts. When some other attacking
scenarios are available, the security of multistep parameter
modulation will be dramatically downgraded.

Now, let us consider the security against known-plaintext and
chosen-plaintext attacks, where the attacker can get or choose
some plaintexts to carry out the attacks. Such attacks are
feasible in some real applications and become more and more common
in the digital networked world today. In known/chosen-plaintext
attacks, it is obvious that the knowledge about some plaintexts
means the knowledge about the bit assignment of the $6n$ strips:
when $m(t)=0$ (or 1), one immediately knows that the strip on
which a point $(A_m,B_m)$ lies corresponds to a 0-bit (or 1-bit),
and then knows that other two strips marked with the same value of
$b_s$ also correspond to 0-bits (or 1-bits). That is, he can
assign a 0-bit (or 1-bit) to the value of $b_s$ corresponding to
the distinguished strip. Once $n$ 0-bits (or 1-bits) have been
assigned to $n$ different values of $b_s$, the attacker can
directly assign 1-bits (or 0-bits) to all other undetermined
values so as to complete the attack. For the number of required
known/chosen plain-bits in the above attack, we have the following
theoretical result.
\begin{theorem}
Assume that $b_s$ distributes uniformly over the set of $2n$
values and that any two values of $b_s$ are independent of each
other. Then, the average number of required known/chosen
plain-bits in the above known/chosen-plaintext attack is $3n$.
\end{theorem}
\begin{proof}
Denote the $k(\geq 1)$ known/chosen plain-bits by
$B_1,\cdots,B_k\in\{0,1\}$, and the corresponding values of $b_s$
by $b_s^{(1)},\cdots,b_s^{(k)}$. The condition that the attack is
completed for the $k$ known/chosen plain-bits equals to the
following term: $n-1$ values corresponding to 0-bits (or 1-bits)
have occurred in $b_s^{(1)},\cdots,b_s^{(k-1)}$, and $b_s^{(k)}$
is the first occurrence of the last value. Considering that each
value occurs with a uniform probability $p=\frac{1}{2n}$ and any
two values are independent of each other, it is easy to get the
probability that the attack stops with $k$ known/chosen
plain-bits, $P(k)$, as follows:
\begin{equation}
P(k)=\begin{cases}
0, & k<n\\
p(1-p)^{k-n}, & k\geq n.
\end{cases}
\end{equation}
Substituting $k'=k-n$ into the above equation, one can get
$P(k')=p(1-p)^{k'},\forall k'\geq 0$. It is obvious that $P(k')$
obeys a geometric distribution, and one can immediately deduce
that $E(k')=p^{-1}=2n$ \citep{GeometricDistribution}. That is,
$E(k)=E(k'+n)=E(k')+n=3n$. The proof is thus completed.
\end{proof}

\begin{figure}[!htb]
\centering
\begin{minipage}{\figwidth}
\centering
\begin{overpic}[width=\textwidth]{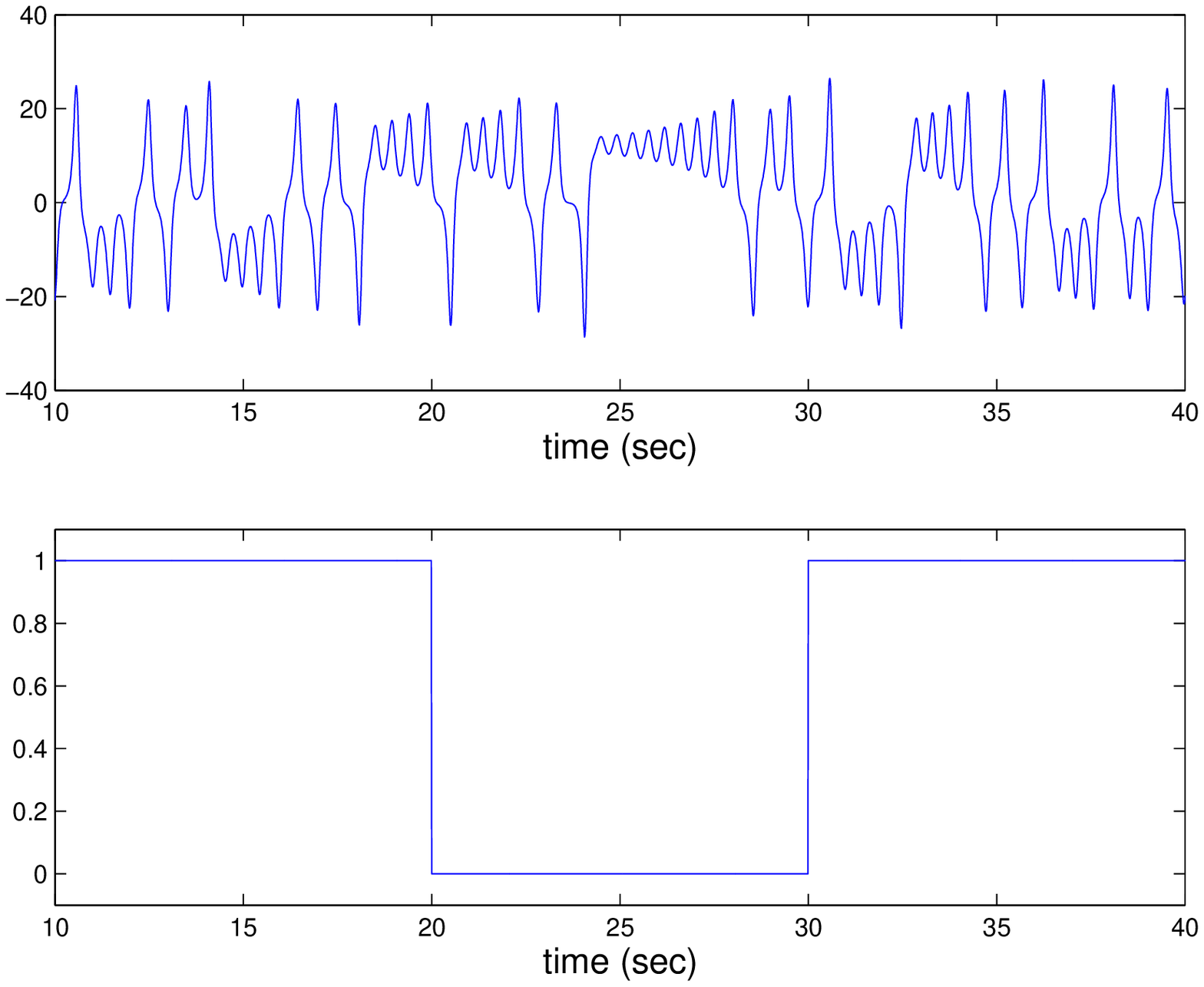}
    \put(10,52){$x_s(t)$}
    \put(10,10){$m(t)$}
\end{overpic}
a) $x_s(t)$ vs $m(t)$
\end{minipage}
\begin{minipage}{\figwidth}
\centering
\includegraphics[width=\textwidth]{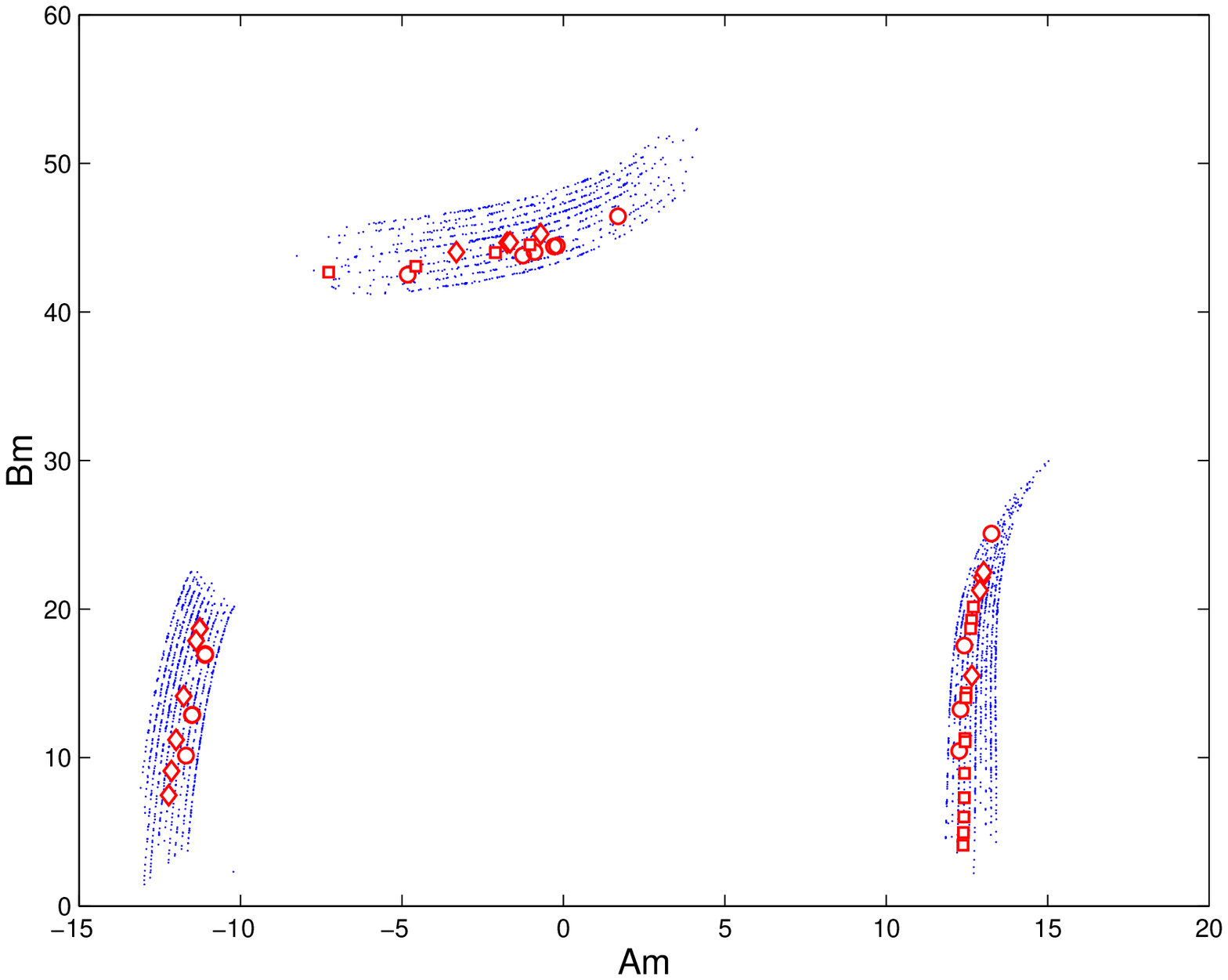}
b) the points $(A_m,B_m)$ vs the return map
\end{minipage}
\begin{minipage}{\figwidth}
\centering
\includegraphics[width=\textwidth]{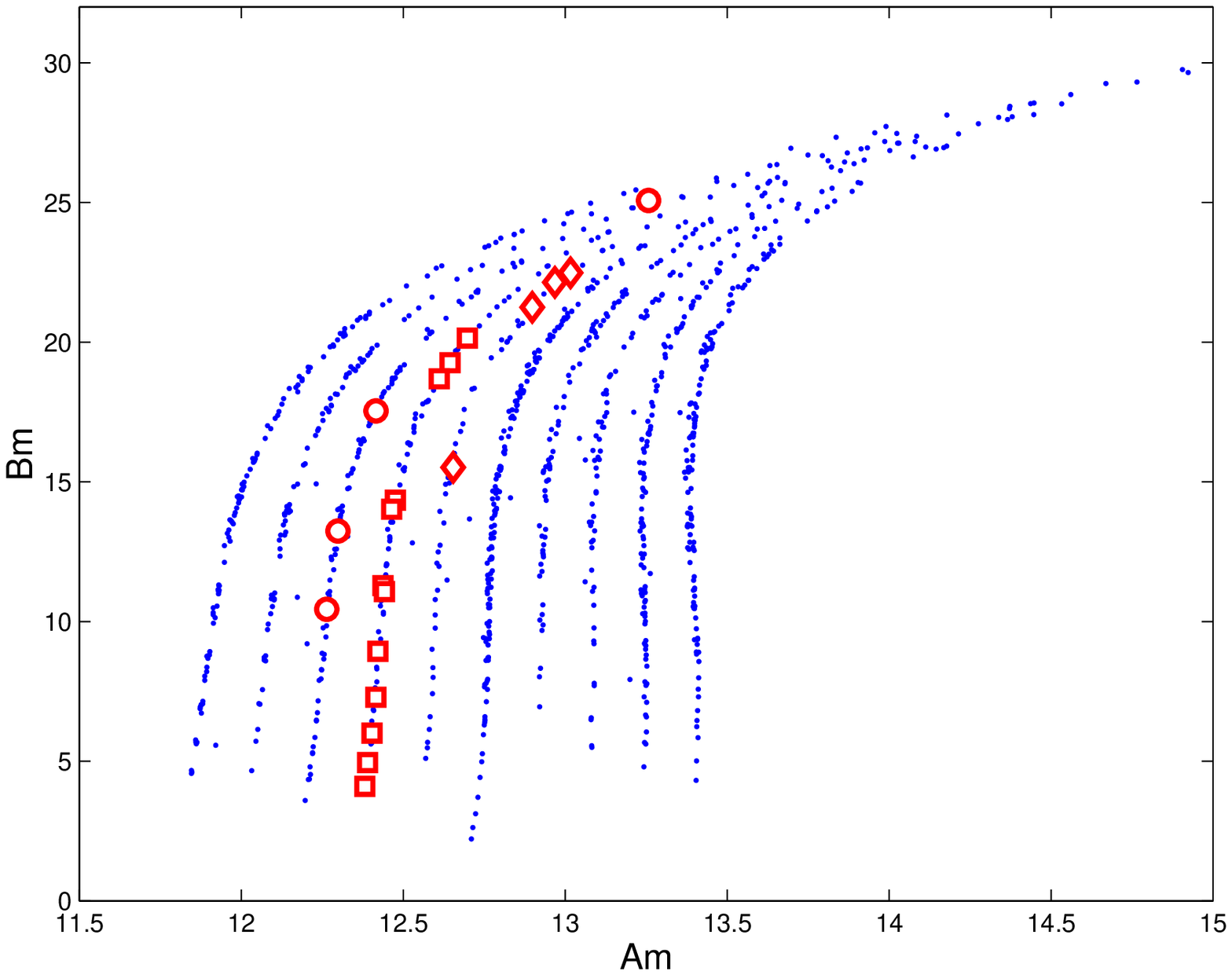}
c) $(A_m,B_m)$ vs the return map: Segment 2
\end{minipage}
\begin{minipage}{\figwidth}
\centering
\includegraphics[width=\textwidth]{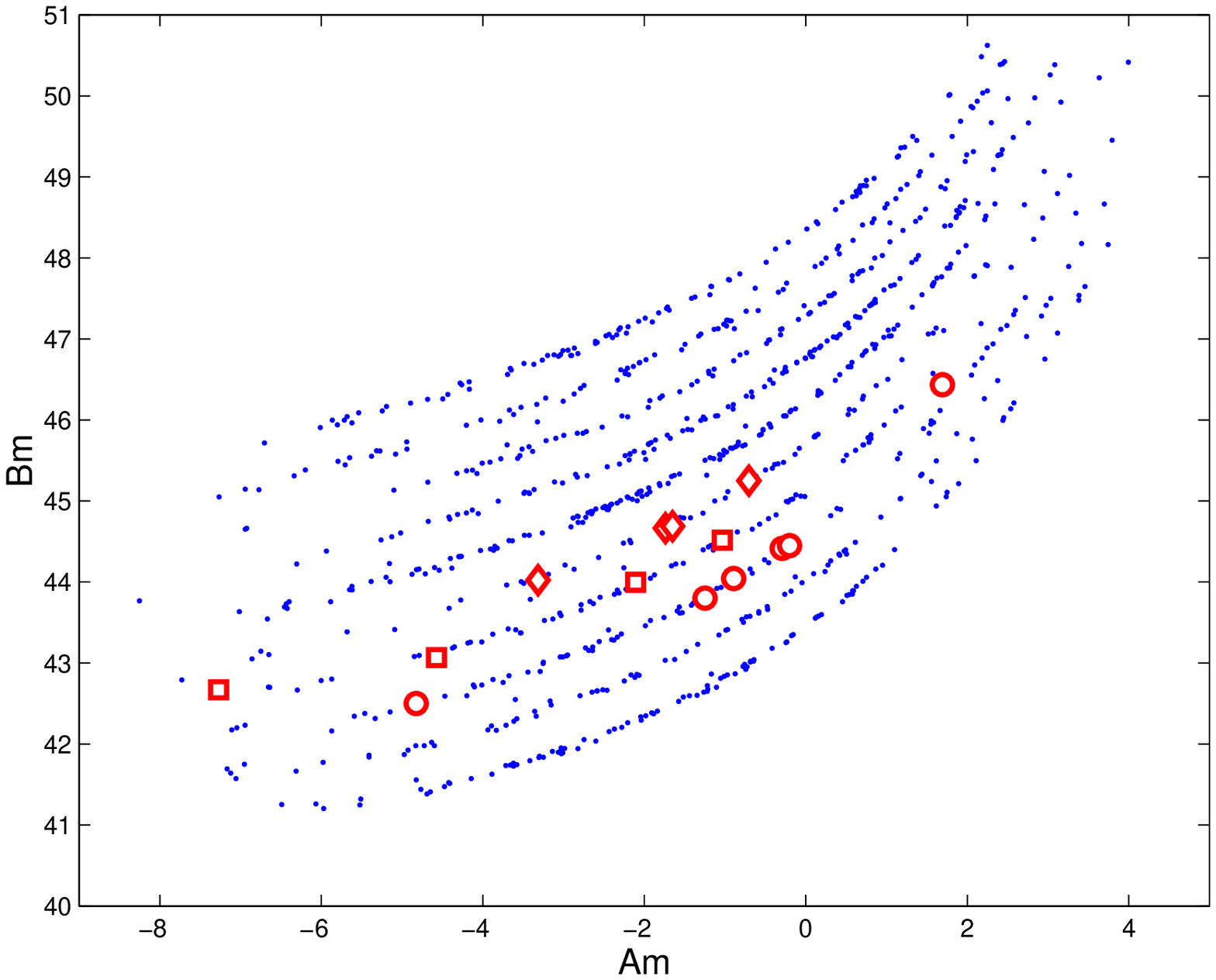}
d) $(A_m,B_m)$ vs the return map: Segment 3
\end{minipage}
\caption{The known/chosen-plaintext attack to multistep parameter
modulation, when $10\leq t\leq 30$. Legend:
\textcolor{red}{$\Diamond$} -- $0\leq t\leq 10$, $m(t)=1$,
$b_s=3.5$; \textcolor{red}{$\ocircle$} -- $10\leq t\leq 20$,
$m(t)=1$, $b_s=3.3$; \textcolor{red}{$\Box$} -- $20\leq t\leq 30$,
$m(t)=0$, $b_s=3.4$.}\label{figure:RM_Attack_MSPM}
\end{figure}

Since $n$ cannot be too large to make the cryptosystem practical
in real applications, the above theorem shows that multistep
parameter modulation is not sufficiently secure against
known/chosen-plaintext attacks. In Fig.
\ref{figure:RM_Attack_MSPM}, we give an example of
known/chosen-plaintext attacks. It can be seen that three
different values of $b_s$, i.e., nine strips in the return map,
are successfully distinguished with only three known/chosen
plain-bits.

\psfrag{time (sec)}{time (sec)}

\section{Breaking Alternative Driving of Transmitter Variables}
\label{section:BreakingADTV}

In this section, we consider how to break another countermeasure
-- alternative driving of transmitter variables. Following the
example given in \citep{Indian:MSCPM:IJBC2001}, we focus on the
$x$/$y$-driving of the Lorenz system. Although the alternative
driving can make the return map less clearer by introducing
overlaps of the $x_s$-map and the $y_s$-map, it is found that the
two overlapped sub-maps can be easily separated so that an attack
can be carried out on the two sub-maps separately.

\begin{figure}[!htb]
\centering
\includegraphics[width=0.7\textwidth]{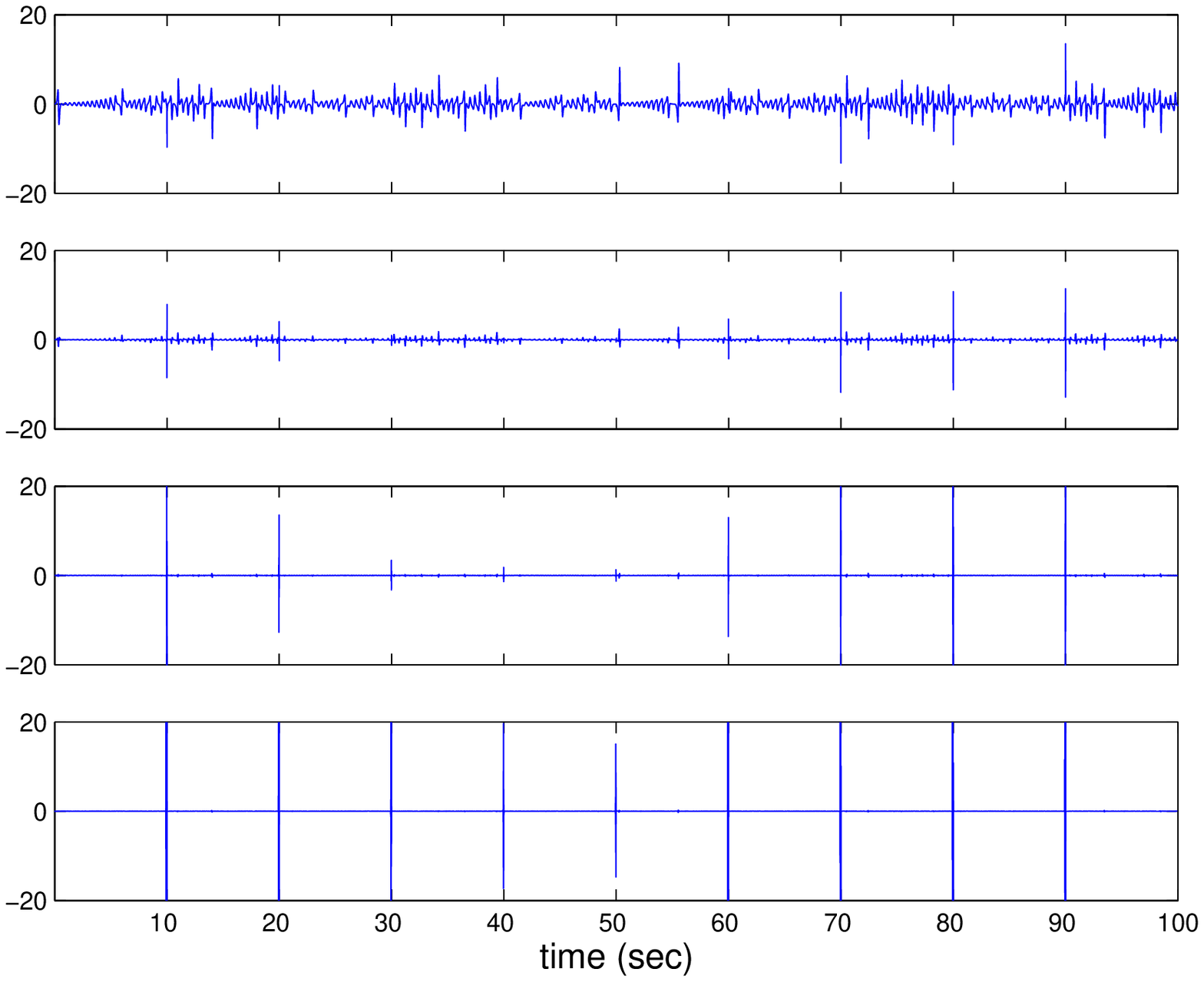}
\caption{The first-order, second-order, 4th-order and 8-th order
(from top to bottom) discrete-time differentiations of the
transmitted signal $d_s$, where $\Delta
t=0.01$.}\label{figure:differences_ds}
\end{figure}

Since there are only two possible driving signals, the separation
of the two driving signals can be simplified to the problem of
detecting the times at which the driving signal, denoted by $d_s$
here, changes from $x_s$ to $y_s$ or from $y_s$ to $x_s$. This can
be easily done by observing the differentiations of $d_s$, since
the alternative driving will introduce breaking points at each
switching time (i.e., discontinuities in $d_s$). Considering that
chaotic signals $x_s(t)$ and $y_s(t)$ are both continuous, the
switching times can be easily distinguished from sudden and large
differentiations of $d_s$, where the word ``sudden" means that the
differentiation at a time $t$ is much larger than the others
around it. In Fig. \ref{figure:differences_ds}, the first-order,
second-order, 4th-order and 8th-order discrete-time
differentiations of $d_s$ are shown for demonstration, where the
display range on the $y$-axis is always limited within $[-20,20]$
to emphasize some sudden and large changes of differentiations
with relatively small amplitudes. It can be seen that all
switching times are sufficiently prominent in the 8th-order
differentiations. Once the switching times are detected, one can
easily separate the $x_s$-map and the $y_s$-map to break the
multistep parameter modulation as discussed in the last section.

\begin{figure}[!htb]
\centering
\psfrag{error point}{error point}%
\psfrag{error points}{error points}%
\psfrag{ambiguous point}{ambiguous point}%
\begin{minipage}{\figwidth}
\centering
\begin{overpic}[width=\textwidth]{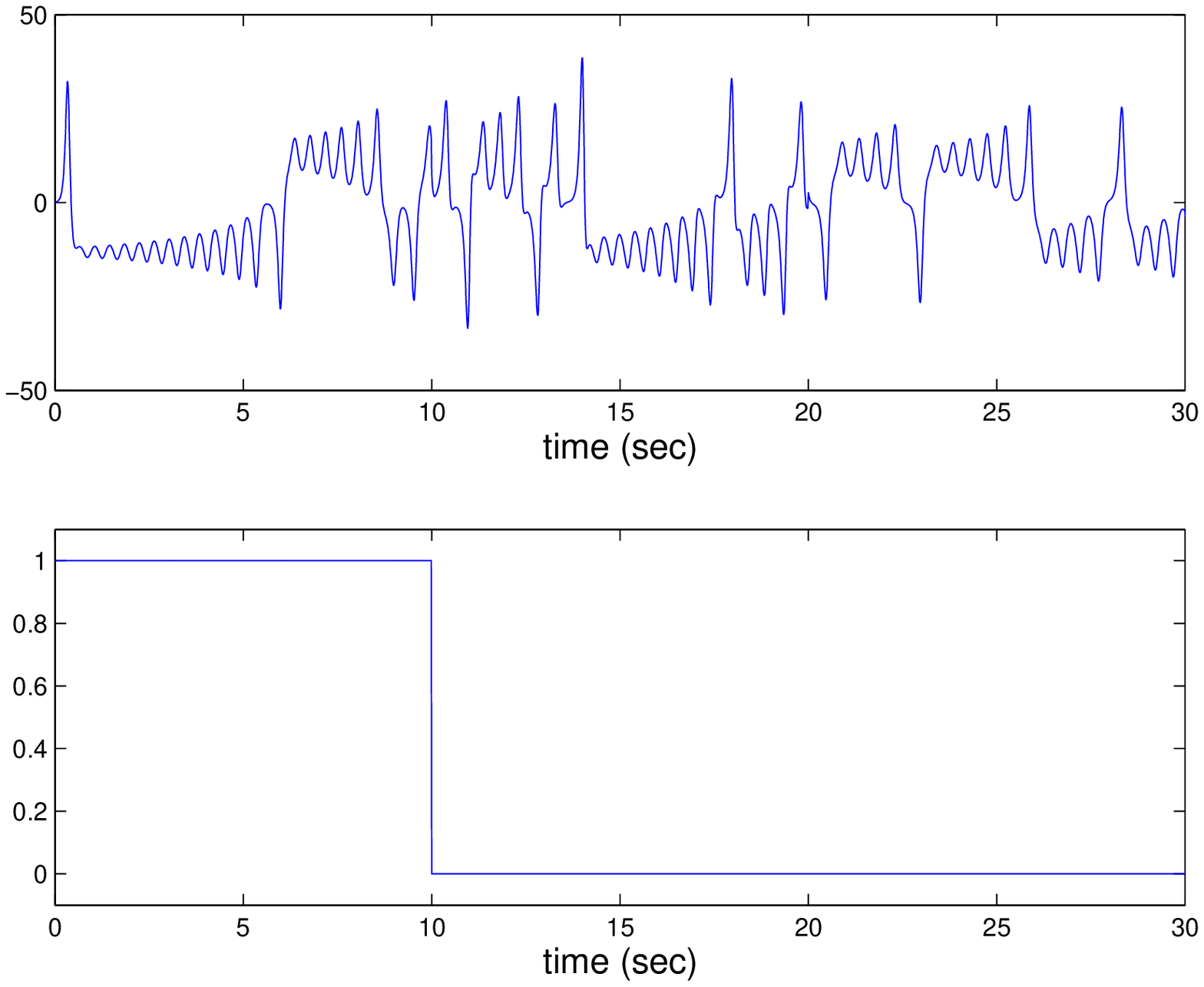}
    \put(10,52){$x_s(t)$}
    \put(10,10){$m(t)$}
\end{overpic}
a) $x_s(t)$ vs $m(t)$
\end{minipage}
\begin{minipage}{\figwidth}
\centering
\includegraphics[width=\textwidth]{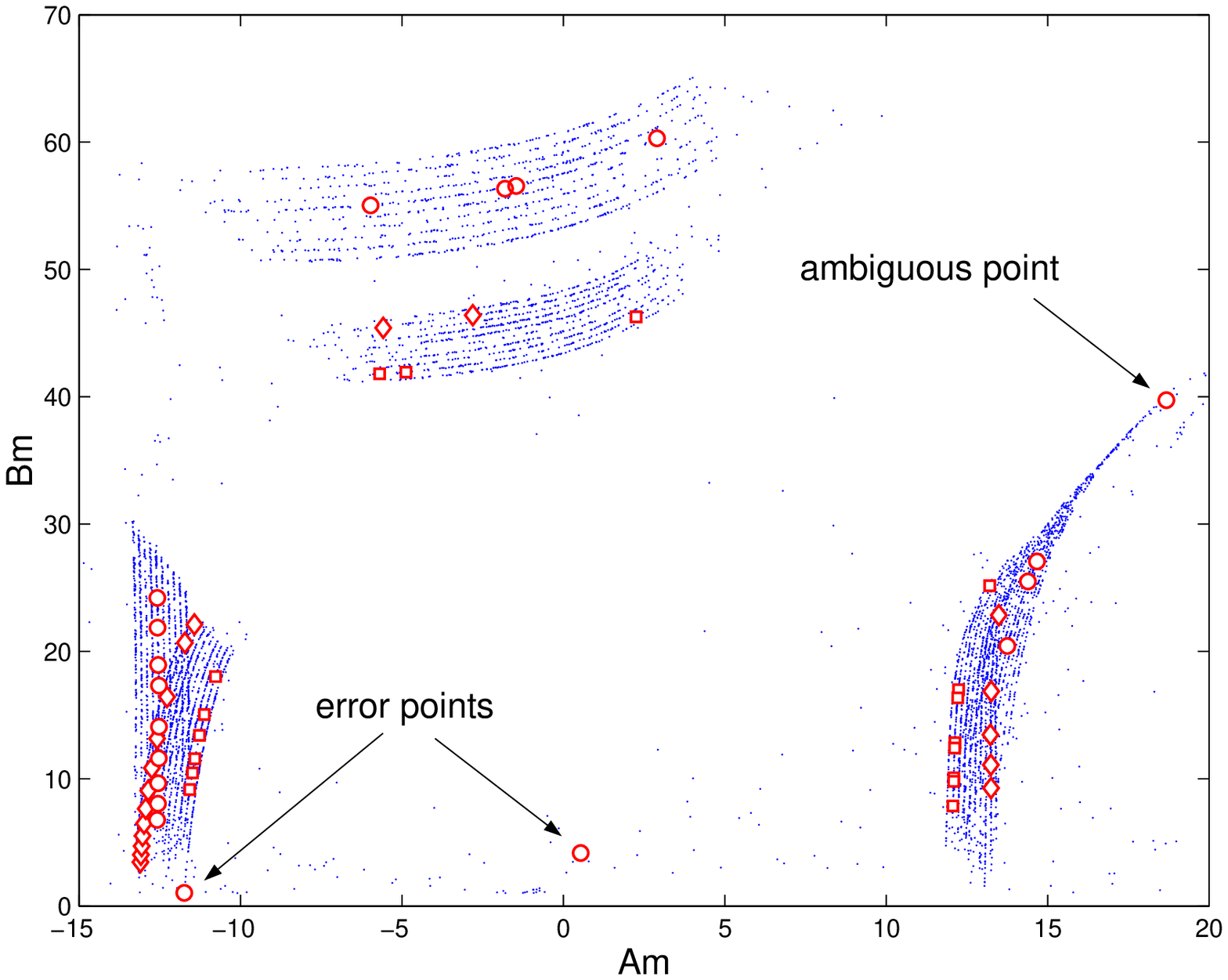}
b) the points $(A_m,B_m)$ vs the return map
\end{minipage}
\begin{minipage}{\figwidth}
\centering
\includegraphics[width=\textwidth]{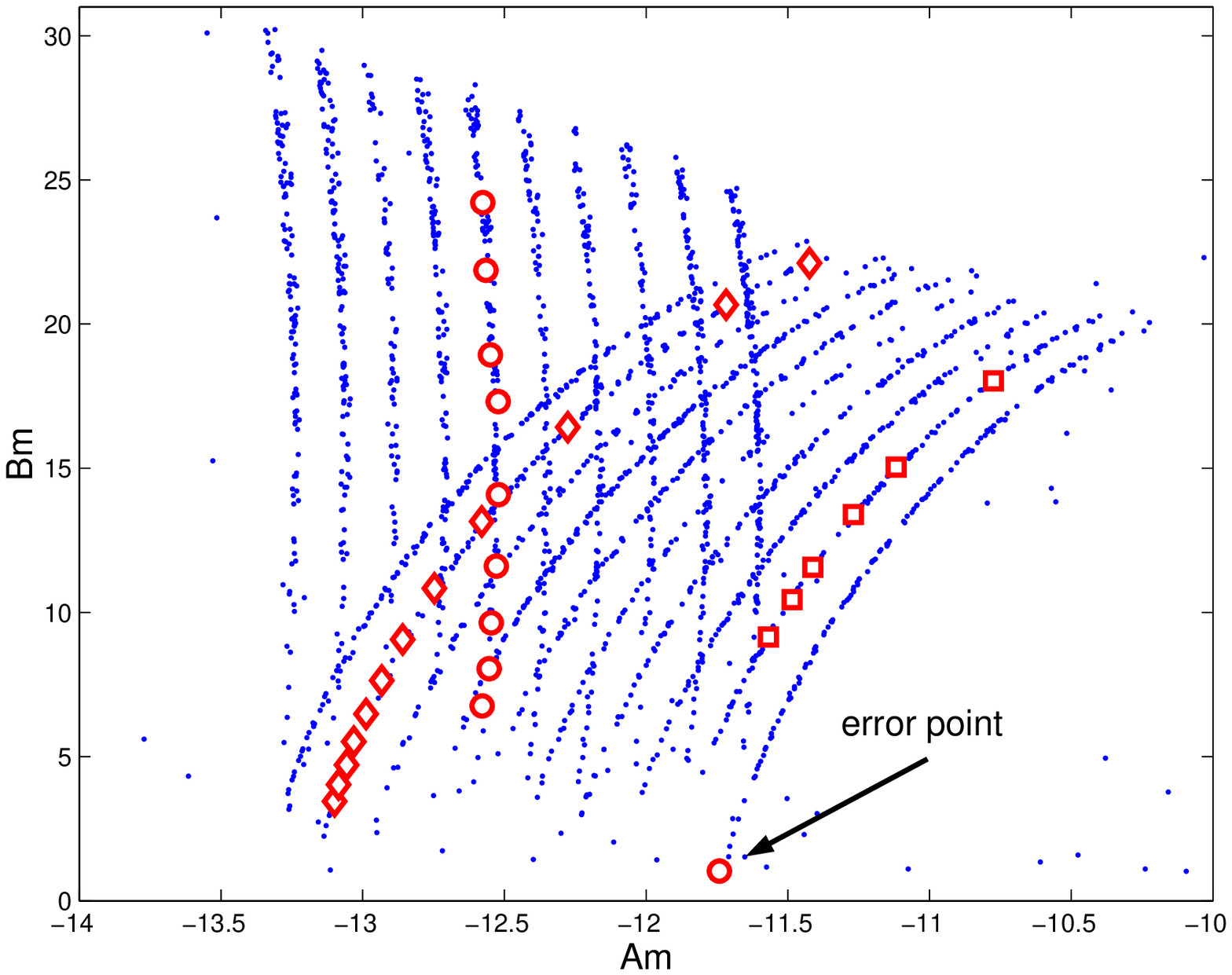}
c) $(A_m,B_m)$ vs the return map: Segment 1
\end{minipage}
\begin{minipage}{\figwidth}
\centering
\includegraphics[width=\textwidth]{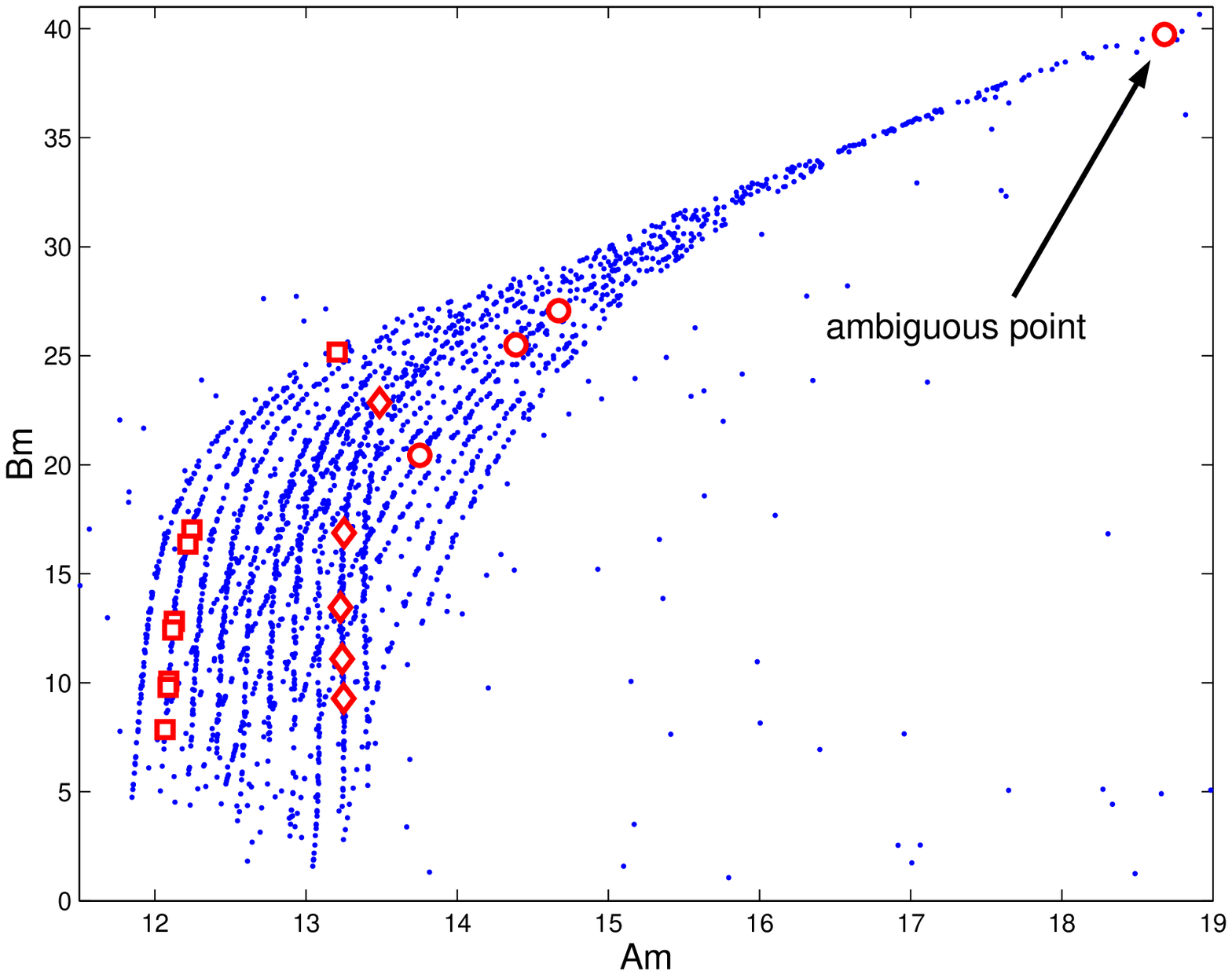}
d) $(A_m,B_m)$ vs the return map: Segment 2
\end{minipage}
\caption{The known/chosen-plaintext attack to multistep parameter
modulation, when $10\leq t\leq 30$. Legend:
\textcolor{red}{$\Diamond$} -- $0\leq t\leq 10$, $x_s$-driving,
$m(t)=1$, $b_s=3.9$; \textcolor{red}{$\ocircle$} -- $10\leq t\leq
20$, $y_s$-driving, $m(t)=0$, $b_s=3.6$; \textcolor{red}{$\Box$}
-- $20\leq t\leq 30$, $x_s$-driving, $m(t)=0$,
$b_s=3.2$.}\label{figure:RM_Attack_MSPMAD}
\end{figure}

In fact, it is even possible to directly separate the two sub-maps
without calculating differentiations of $d_s$. Observing Fig.
\ref{figure:RM_MSPMAD}, one can find that the overlaps of the two
sub-maps are not very significant, which makes it possible to
separate the two sub-maps directly from the alignment directions
of consecutive points $(A_m,B_m)$. When $x_s$-driving is used for
odd bits and $y_s$ for even bits,
Fig.~\ref{figure:RM_Attack_MSPMAD} shows the positions of the
points $(A_m,B_m)$ in the return map for $0\leq t\leq 30$. In
spite of the existence of a few error points and ambiguous points,
which are mainly introduced by the faked maxima and minima near
the switching times, it is still very easy to distinguish which
driving signal was used from the alignment direction of the points
$(A_m,B_m)$ corresponding to the current bit (i.e., to the current
value of $b_s$). The accidental errors and ambiguous points can be
easily removed by filtering techniques.

Finally, we examine the attack complexity when both
countermeasures are used in a secure communication system. Since
there exist $12n$ strips, the average number of plain-bits in
known/chosen-plaintexts attacks will be $2\cdot 3n=6n$, which
means that the security against known/chosen-plaintext attacks is
still rather weak. The security against ciphertext-only attacks is
relatively higher: $\left(2\cdot\binom{2n}{n}\right)^2$. However,
note that an attacker can extract 50\% of all plain-bits, even
when he only exhaustively guesses the right bit assignment
corresponding to the $x_s$-map or the $y_s$-map. Thus, strictly
speaking, the security against ciphertext-only attacks is still in
the order of $2\cdot\binom{2n}{n}$, i.e., the same as the one
under the condition that only the first countermeasure is used. As
mentioned above, to make the designed secure communication system
sufficiently secure, $n\geq 50$ is required.

\section{Conclusion}

To resist the return-map attack presented in
\citep{Perez:ReturnMapCryptanalysis:PRL95},
\cite{Indian:MSCPM:IJBC2001} proposed two countermeasures to
enhance the security of the chaotic switching (i.e., binary
parameter modulation) scheme. After refining the return-map attack
by exploiting a deterministic relationship between the return map
and the modulated parameter, this paper points out that these two
countermeasures are not secure enough against
known/chosen-plaintext attacks. Also, it is found that the
security against ciphertext-only attacks cannot be ensured if the
proposed secure communication system contains less than 200
sub-systems.

The cryptanalysis results given in this paper show that one has to
use more powerful techniques to effectively resist return-map
attacks. Recently, a new CSK scheme was proposed in
\citep{XuChee:CSKwFS:IJBC2004} by introducing many false switching
events. It is under study whether or not this new CSK scheme is
secure against the return-map attack described in this paper. At
present, it is still an open problem to design a chaos-based
secure communication system that is strong enough against all
known attacks, and to find more powerful cryptanalysis tools to
evaluate the security of various chaos-based cryptosystems.

\section*{Acknowledgements}

This research was partially supported by the Applied R\&D Centers
of the City University of Hong Kong under grants no. 9410011 and
no. 9620004, and by the Ministerio de Ciencia y Tecnolog\'{\i}a of
Spain, under research grants TIC2001-0586 and SEG2004-02418.

\end{document}